\documentclass[sn-mathphys]{sn-jnl}
\jyear{2021}
\theoremstyle{thmstylethree}
\newtheorem{theorem}{Theorem}

\theoremstyle{thmstyletwo}

\theoremstyle{thmstylethree}

\usepackage{ragged2e}
\usepackage{scrextend}
\usepackage{inputenc}
\usepackage{titlesec}
\usepackage{algpseudocode}
\usepackage{algorithm}
\usepackage{hyperref}
\usepackage[symbol]{footmisc}

\begin{document}

\title[{\footnotesize PiouCrypt: Decentralized Lattice-based Method for Visual Symmetric Cryptography}]{PiouCrypt\footnote[2]
	{This project's source code is available at \href{https://bitbucket.org/my_piou/PiouCrypt}{https://bitbucket.org/my_piou/PiouCrypt}.}
	: Decentralized Lattice-based Method for Visual Symmetric Cryptography\footnote[4]{This is an earlier draft of the manuscript (Last Updated: April, 2022) \\ Written in Springer Nature \LaTeX Template}}

\author[1]{\fnm{Navid} \sur{Abapour$^{*}$}}

\author[2]{\fnm{Mohsen} \sur{Ebadpour}}

\affil[1]{\orgdiv{{\small Department of Computer Science}}, \orgname{{\small University of Mohaghegh Ardabili}}}
\affil[2]{\orgdiv{{\small Department of Computer Engineering}}, \orgname{{\small Amirkabir University of Technology}}}

\affil[*]{Corresponding Author: \href{navidabapour@gmail.com}{mailto:navidabapour@gmail.com}}

\abstract{	
			In recent years, establishing secure visual communications has turned into one of the essential problems for security engineers and researchers. However, only limited novel solutions are provided for image encryption, and limiting the visual cryptography to only limited schemes can bring up negative consequences, especially with emerging quantum computational systems. This paper presents a novel algorithm for establishing secure private visual communication. The proposed method has a layered architecture with several cohesive components, and corresponded with an NP-hard problem, despite its symmetric structure. This two-step technique is not limited to gray-scale pictures, and furthermore, utilizing a lattice structure causes to proposed method has optimal resistance for the post-quantum era, and is relatively secure from the theoretical dimension.
		 }


\maketitle

\section{Introduction}\label{sec1}

Due to the rapid expansion of the Internet and its features such as social networks \cite{bib1}, network scales with specific protocols \cite{bib2}, secure cloud memory \cite{bib3}, and smart contracts \cite{bib4} on the one hand, and the rise in people's privacy concerns on the other, transferring and storing images in a secure manner is a critical point \cite{bib5, bib6}. Encrypting images with well-known algorithms like DES or AES may appear to be a solution at first glance, but it does not meet all of the requirements. The fundamental reason is that the aforementioned algorithms encrypt data regardless of the strong correlations between data which is a crucial aspect of adjacent pixels of images. Furthermore, they do not take into account the characteristics of the high redundancy in images. As a result, providing a mechanism that efficiently encrypts this type of data is crucial \cite{bib7}.

Nowadays, a diverse set of solutions (which some of them mentioned in Section \ref{sec2}) as mechanisms are getting provided with more weights on both image processing \cite{bib8, bib9} and cryptographic aspects \cite{bib10}. Also, due to advances of computational models in recent years, creating cryptosystem based on lattice structures is getting more common because of their potential for bearing expectations in post-quantum era \cite{bib11}; not only limited to high-level primitives, but even in authentication procedure and zero-knowledge-based protocols \cite{bib12}. Another rare and novel solution can be using privacy preserving models for generalizing visual cryptosystems to multi-party computations \cite{bib13}, or moving deeper levels of pixels and applying disjunctive normal form or computational logic-based approaches \cite{bib14}. Creating techniques for visual cryptosystems that are circulating around random oracle models was another solution in the past, but currently it can not be applicable according to these models limitations \cite{bib15}.

Another solution, entitled "PiouCrypt", is proposed in this work, which is a novel security architecture for visual communications. PiouCrypt is transferring one key between the sender and receiver with a stream symmetric scheme because block ciphers in visual models, specifically in PiouCrypt, can be really vulnerable to attacks like overrefined system of equations \cite{bib16}. The decentralized structure of PiouCrypt uses several components to achieve a secure method. For example, the non-negative matrix factorization is an NP-hard problem (based on Vavasis's theorem \cite{bib17}) that despite of piouCrypt's symmetric scheme, it is used to decrease the computational vulnerabilities.

This paper, apart from current section, is divided into six parts: Section \ref{sec2} circulated around the related precedent of visual cryptography. The \hyperref[sec3]{third} section presents the idea of PiouCrypt with details. Part \ref{sec4} takes a quick look at the security analysis of PiouCrypt from the point of image processing view. The \hyperref[sec5]{fifth} part compares the pros and cons of related works with the proposed technique. Section \ref{sec6} talks about the potential of PiouCrypt for further developments in the future and a conclusion of work is drawn in Part \ref{sec7}.

\section{Background and Related Works}\label{sec2}
Although the limitations of applicable general ideas in image encryption, from chaotic aspect there are many tools which are common in modern visual cryptography: Coupled Map Lattice \cite{bib18}, Piece-wise Logistic Map \cite{bib19}, Partioned Cellular Automata \cite{bib20}, Keyed Hash Functions \cite{bib21}, Nonlinear Filtering Function \cite{bib22}, Baker Map Scrambling \cite{bib23}, Self-synchronizing Steam Ciphers \cite{bib24}, Non-adjacent Coupled Map Lattices \cite{bib25}, Logistic-dynamic Arnold Lattice Model \cite{bib26}, Spatiotemporal Chaos-coupled Coefficient Model \cite{bib27}. \par{}

Chaotic and permutation-based techniques are mostly used in recent researches because of their strong mathematical structures (ergodicity and exactness), and butterfly effects. Despite the fact that there are methods with two layer for encrypting image \cite{bib28}, PiouCrypt has two separated and decentralized layer to secure the visual communication. In the continue, three different techniques are discussed as related works, and in the Section \ref{sec5} they are getting compared with our proposed method.

In \cite{bib26}, a technique was presented based on non-adjacent coupled map lattices, where the initial configuration was going to set by the equation \ref{eq-A}, by assuming every $k_{i}$ is single block of the external hash key $K$, and $t_{i}$ representing values set by developer such that $i={1,2,...,8}$. The space of key is approximately $S_{k e y}=10^{14 \times 8} \times 2^{256} \approx 2^{656}$.

\begin{align}
	\left\{\begin{array}{l}
	t_{i}^{\prime}=t_{i}+\frac{\left(k_{i} \oplus k_{i+8} \oplus k_{i+16}\right)+t(i)}{(k+t) \times 10^{3}} \\
	t=\frac{\operatorname{sum}\left(\bmod \left(t_{i} \times 10^{14}\right), 256\right)}{8}
	\end{array}\right.
	\label{eq-A}
\end{align}

After initialization with matrix of Latin square, every pixel of the input picture was getting encrypted only once, and by decreasing multiple rounds of encryption, it was trying to avoid complexity. Based on predefined equations, some kind of diffusion was created in the image, then the permutation was applying to image's channels, and after the switch of value between them and single channel, a substitution process starts to implement substitute pixel's S-box to modify them for preparing cipher image.

Another related work was proposed in \cite{bib29} based on Arnold logistic lattice model, where the key similar to method of \cite{bib26} was separating into several parts, as well as input image. Next, the value of key gets involved with a logistic-dynamic coupled map lattice model (LDCML) as equation \ref{eq-B} shows; where logistic map denoted by $f(x)$, and sequence number and iterations are $i$ ($L \geq i \geq 1$) and $n$ respectively. Coupling coefficient is shown by $e$, and multiple parameters considered fixed by developers.

\begin{equation}
	\begin{aligned}
		&{\left[\begin{array}{l}
				j \\
				k
			\end{array}\right]=\left[\begin{array}{cc}
				1 & p \\
				q & p q+1
			\end{array}\right]\left[\begin{array}{l}
				i \\
				i
			\end{array}\right]} \\
		&(\bmod L)(\text { if } j==0, j=L . \text { if } k==0, k=L) \\
		&f\left(x_{n}\right)=\mu x_{n}\left(1-x_{n}\right) \\
		&e_{n+1}=\mu_{1} e_{n}\left(1-e_{n}\right)\left(\mu_{1} \in(3.57,4], e_{1} \in[0,1]\right) \\
		&x_{n+1}(i)=\left(1-e_{n}\right) f\left(x_{n}(i)\right) \\
		&\quad+\left(e_{n} / 2\right)\left(f\left(x_{n}(j)\right)+f\left(x_{n}(k)\right]\right)	
	\end{aligned}
		\label{eq-B}
\end{equation}

After repeating a permutation procedure, eight groups of $Z_{i}$ is going to generated by an special grouping method, and values of the first five lattices of the last iteration are going to enter a predefined model to modify the pixel stream. The cipher image will be ready if the reshaped version's rounds of final pixel stream are fulfilling some computational circumstances. Also, this work provides a bidirectional diffusion technique based on twice independent. Similar to \cite{bib26} and \cite{bib29}, in \cite{bib30} another chaos based method presented using chaotic spatiotemporal model. However, in \cite{bib30}, the algorithm divides key into twelve groups, and uses a LDCML system like Ref. \cite{bib29} for iteration. After pixel level permutations, the plain image gets separated and combined with permutation and iterations on logistic map. Like Ref. \cite{bib29}, the cipher image is going to produced based on the encryption rounds and key's cyclic shifts.

\section{Proposed Technique}\label{sec3}
This technique consists of two layers for improving the security level; the first layer is structured by computational and algebraic functions, and the second layer is a simple but effective step based on American Standard Code for Information Interchange (ASCII) encoding.\par
	\begin{figure}[h]%
		\centering
		\includegraphics[trim={0 0 1mm 0},clip, width=1\textwidth]{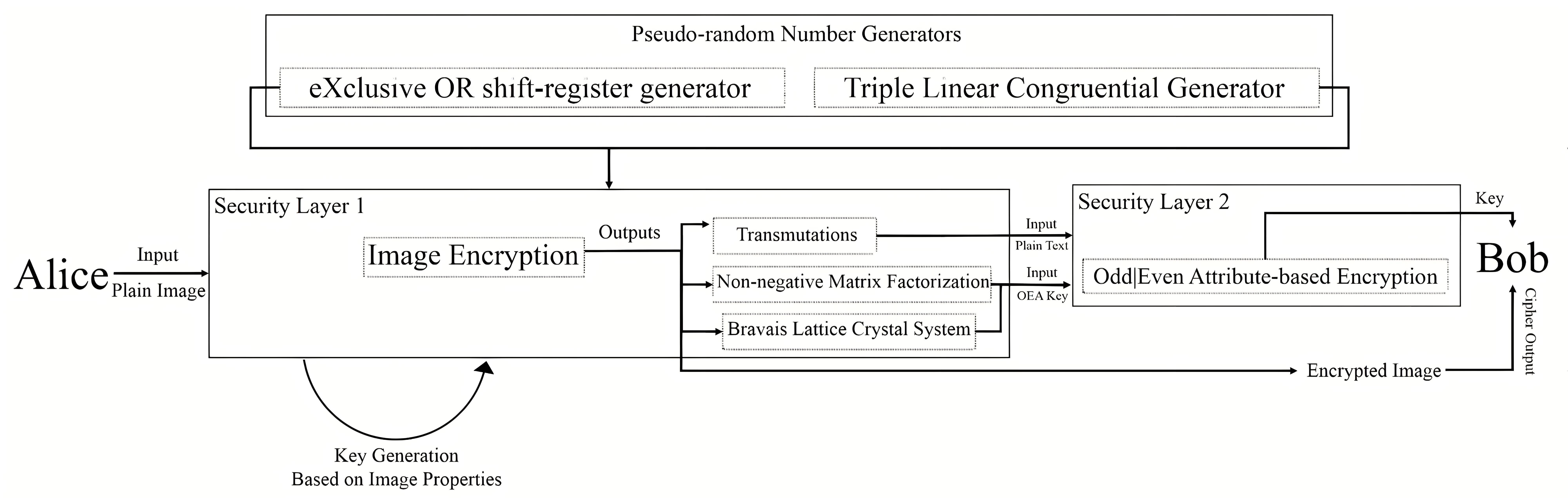}
		\caption{General Presentation of Proposed Method}\label{fig-1}
	\end{figure}
Initially, Alice sends her image as an input, and the first security layer starts it work by applying multiple modifications on picture. Then, based on the data gathered from plain image, a Bravais lattice will generated, and it is going to get mapped into a 2D non-negative matrix, which it's factorization will get calculated. Afterward, the information produced in first layer will transmitted to the second layer of method. This step will encrypt the information based on numbers parity attribute by using several entangled operations. Eventually, the generated key in Security Layer 2 most deliver to Bob, along with encrypted image from Security Layer 1 by Alice.

\subsection{Primary Security Layer}\label{subsec3_1}
The first layer decomposes the plain image into three channels of Red, Green, and Blue (RGB), and uses XoRShift Pseudo-random Number Generator (PRNG) to generate two values $i$ and $j$. In every RGB channel the $i^{th}$ row will get swapped with $j^{th}$ row, and the $i^{th}$ column will get swapped with $j^{th}$ column, and this procedure will continue according to width or length of input picture (whichever is greater). The values of $i$ and $j$ are going to stored (e.g. in a file named "\textit{Layer\_1-Keys}") in order to use them in decryption. After making the plain image messy by creating noises, all three RGB channels will get merged to construct the cipher image. Meanwhile, Naor–Reingold Pseudo-random Function (NRPF) is going to generate two vectors for forming up the Bravais Lattice Crystal System with well-formed spots in it. The collection of these points will create a two dimensional matrix wit non-negative values. Next, this matrix will get decomposed into two matrices $W$ and $H$. The value of $W$ is going to stored (e.g. in a file named "\textit{OEA\_key}") for utilizing it in Security Layer 2.

\begin{figure}[h]%
	\centering
	\includegraphics[trim={0 0 1mm 0},clip, width=0.9\textwidth]{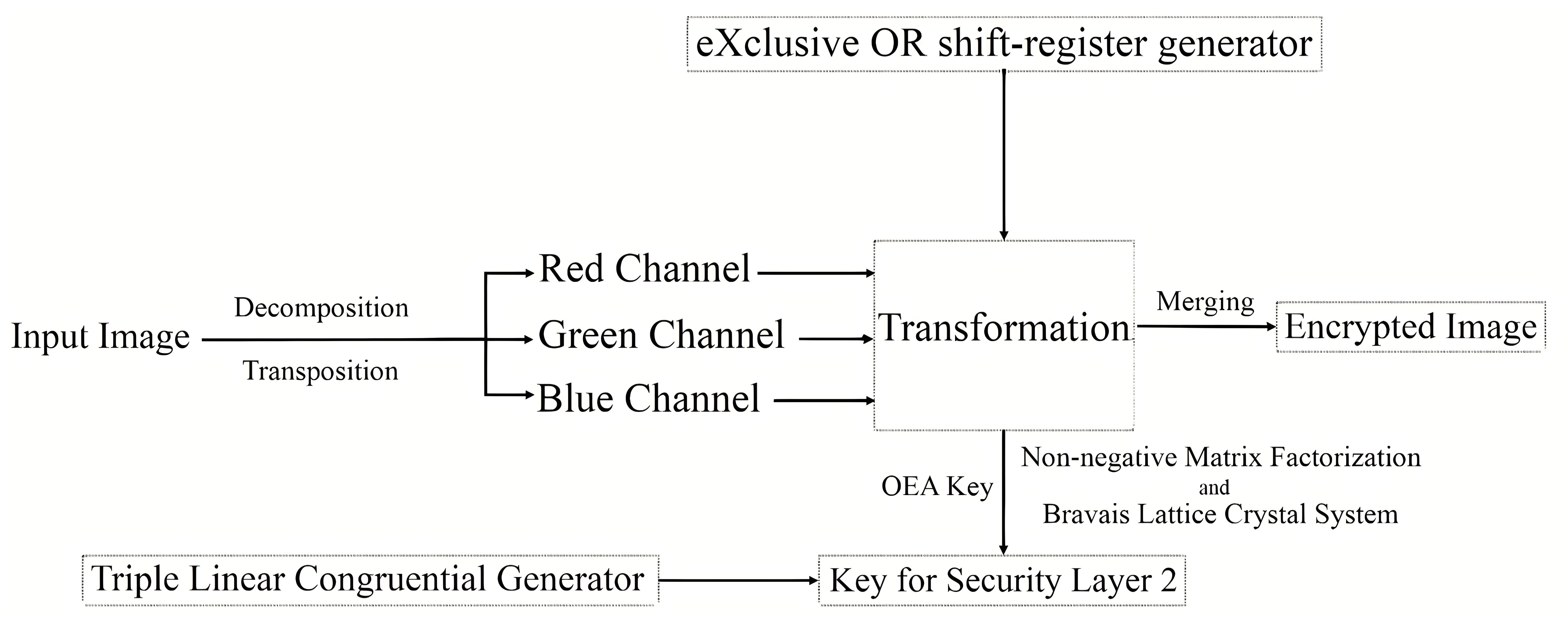}
	\caption{Procedure of Primary Security Layer}\label{fig-2}
\end{figure}

The PiouCrypt is utilizing 64-bit modern implementation of Saito and Matsumoto's XSadd generator \cite{bib31, bib32}: eXclusive OR Shift-register Generator (Xorshift1024$^{*}$). Considering the main focus in on visual massage, the Xorshift generator can be a suitable candidate for getting used here, because the speed of technique can not get ignored \cite{bib33}. A linear representation of this generator can be shown as: \par{}
\vspace{-5mm}
\begin{align}
	\mathbf{v}_{i}=\sum_{j=1}^{r} \mathbf{A}_{j} \mathbf{v}_{i-j} \bmod 2 \ \text { where } \mathbf{A}_{j}=\sum_{\left\{l: m_{l}=j\right\}} \tilde{\mathbf{A}}_{l} \quad \ni 
	\nonumber
	\\
	p,m_{j} \in \mathbb{N} \quad ,\quad  r = max_{1 \geq j \geq p} m_{j}
	\label{eq-1}
\end{align}

The value of $v_{i}$ represents $w$-bit block vectors (where $w$= 32 or 64), and $\tilde{\mathbf{A}_{j}}$ is the product of Xorshift matrices for some $v_{j} \geq 0$. Since there are multiple versions of Xorshift for generating bit-vectors x, the PiouCrypt is using Marsaglia's method: \par{}

\begin{align}
	(x^{(k-r+1)} \mid x^{(k-r+2)} \mid \cdots \mid x^{(k)} )=(x^{(k-r)} \mid x^{(k-r+1)} \mid \cdots \mid x^{(k-1)}) \, \mathcal{C}
	\nonumber
	\\
	s.t. \quad
	x \in F_{2}^{1 \times w} \quad , \quad r>0
	\quad , \quad n=rw \, \, \, or \, \, \, n=2^{k} \, (12 \geq k \geq 6)
	\label{eq-2}
\end{align}

Frobenius companion matrix $\mathcal{C} \in F_{2}^{n \times n}$ represents the $r \times r$ matrix of $w \times w$ blocks. In this work, the uniform Xorshift is getting used for selecting rows and columns of plain image to swap, and considering the main focus is on a visual content, so the speed is playing an outstanding role, and Xorshift's linear structure helps PiouCrypt to efficiently play this role. Also, the output of Xorshift (method of Marsaglia) will have better values, if developers set long periods ($>10^{1232}$) for RNG; so, in PiouCrypt the Xorshift is generalized by increasing the amount of cycle length.

\begin{figure}[h]%
	\centering
	\begin{program}
		\BEGIN \\ %
		|uint64\_t| \, \, s[16];
		|int| \, \, p;
		|uint64\_t| \, \, mult := |0x106689D45497FDB5|;
		\PROC |uint64\_t| \, \, xorshift1024star(void) \BODY
		|uint64\_t| \, \, s0 := s[p];
		|uint64\_t| \, \, s1 := s[p := (p + 1) \, |\&| \, 15];
		s1 \, \, \hat{}= s1 \ll 31; \rcomment{The operator $\hat{}$ is equivalent to $\oplus$ in C}
		s[p] := s1 \, \, \hat{} \, \, s0 \, \, \hat{} \, \, (s1 \gg 11) \, \, \hat{} \, \, (s0 \gg 30);
		|return| \, \, s[p] * mult; \ENDPROC
		\END
	\end{program}
	\caption{C-like functional definition of Xorshift1024$^{*}$ with a fixed seed}\label{fig-3}
\end{figure}

In PiouCrypt, a modification have been applied to Xorshift, in order to avoid extreme bias based on R.B.Davies's theorem.

\begin{theorem}[Independent Biased Pairs]\label{thm-1}
	Let X, Y denote random bits and assume E(X) as expected value of X (the average value of a large number of repeated trials). If $X$ and $Y$ are independent random bits with $E(X)=\mu$ and $E(Y)=\nu$ then
	\begin{align}
		E(X \otimes Y)=\mu+\nu-2 \mu \nu=\frac{1}{2}-2\left(\mu-\frac{1}{2}\right)\left(\nu-\frac{1}{2}\right)
		\label{eq-3}
	\end{align}
	Assuming $E(X)=E(Y)=\mu$, and independence of $X$ and $Y$:
	\begin{align}
		E(X \otimes Y)=2 \mu(1-\mu)=\frac{1}{2}-2\left(\mu-\frac{1}{2}\right)^{2}
		\label{eq-4}
	\end{align}
\end{theorem}

This means the short distance between $\nu$ and $\mu$, and $\frac{1}{2}$ yields closeness of $E(X \otimes Y)$ to $\frac{1}{2}$; so, in PiouCrypt the values of $\nu$ and $\mu$ for Xorshift component are getting set in a way they keep distance from $\frac{1}{2}$ to statically reduce the deviation of expectation as more as possible. The inverse of Theorem \ref{thm-1} is notable, and a huge deviation of the expectation can occur via even a tiny value of statistical dependence (correlation).

Not only in PiouCrypt, but in many other applications Xorshift can significantly decrease deviation of the expectation from $\frac{1}{2}$.

The next part of PiouCrypt is the first section of method, where the input picture gets modified, and in the following it has been detailed.

\begin{algorithm}
	\caption{Encrypting Image}\label{algo-1}
	\begin{algorithmic}[1]
		\Require Plain Image
		\Ensure Cipher Image
		\State Decomposition of input picture into red, green, and blue channels
		\State $W \Leftarrow$ plain image's width, $h \Leftarrow$ plain image's height
		\State Create $w \times h$ \textbf{integer} matrices $M\_red$, $M\_green$, and $M\_blue$
		\State $(M\_red, M\_green, M\_blue) \Leftarrow$ values of red, green, and blue channels
		\If{$w < h$}
			\State \textit{counter\_row} $\Leftarrow h$
		\Else
			\State \textit{counter\_coloumn} $\Leftarrow w$
		\EndIf
		\While{\textit{counter\_row} $\neq 0$}
			\State $i,j \Leftarrow$ generate two numbers by Xorshift
			\State \textbf{swap} $i^{th}$ row of matrices $M\_red$, $M\_green$, $M\_blue$ with $j^{th}$ row
			\State \textbf{store} $i$ and $j$ in Layer\_1-Keys.txt
			\State \textit{counter\_row} $\Leftarrow$ \textit{counter\_row} $- \, 1$
		\EndWhile
		\While{\textit{counter\_coloumn} $\neq 0$}
		\State $i,j \Leftarrow$ generate two numbers by Xorshift
		\State \textbf{swap} $i^{th}$ column of matrices $M\_red$, $M\_green$, $M\_blue$ with $j^{th}$ column
		\State \textbf{store} $i$ and $j$ in Layer\_1-Keys.txt
		\State \textit{counter\_coloumn} $\Leftarrow$ \textit{counter\_coloumn} $- \, 1$
		\EndWhile
		\State \textbf{merge} ($M\_red$, $M\_green$, $M\_blue$) into channels of a new image 
		\State \textit{counter\_LUT} $\Leftarrow$ $255$ \label{algo-1-23}
		\While{\textit{counter\_LUT} $\geq 0$}
			\State Layer\_1-Keys.txt $\Leftarrow$ \textbf{print} \textit{counter\_LUT}
			\State $z \Leftarrow$ generate a number by Xorshift
			\For{every pixel \textbf{in} image}
				\If{pixel's value == \textit{counter\_LUT}}
					\State pixel's value $\Leftarrow$ $z$
				\EndIf
			\EndFor
			\State Layer\_1-Keys.txt $\Leftarrow$ \textbf{print} \textit{z}
			\State \textit{counter\_LUT} $\Leftarrow$ \textit{counter\_LUT} $- \, 1$
		\EndWhile
		\State Save the cipher image until the procedure of Security Layer 2 complete
	\end{algorithmic}
\end{algorithm}

As it is raised in Algorithm \ref{algo-1}, the plain image will get sloppy, and the information related to this action is going to get saved in a file. However, swapping row and columns in not enough, and the reason is obvious: The adversary can still realize the colors of Alice's picture. Therefore, in \ref{algo-1-23}$^{rd}$ line of Algorithm \ref{algo-1} the PiouCrypt starts to create a lookup table (LUT) for every value of RGB system, in a way that the numbers of from 0 to 255 will get printed in \textit{Layer\_1-Keys.txt}, along with a pseudo-random value in front of each of them. While image encryption, the algorithm will check to see if there is any pixel with the values of lookup table to change it into the pseudo-random value generated by Xorshift. As a result, the whole pixels of plain image will get sloppy and the altered, and attacker will not be able to even realize the original picture's colors.\par{}
The input picture sent by Alice should not deliver to Bob until the completeness of Security Layer 2; also, the \textit{Layer\_1-Keys.txt} file, which the image encrypting data is stored in it, will get transferred to the next section: Bravais Lattice Crystal System. In the following, outputs of different parts of Algorithm \ref{algo-1} is presented in each figure. \par{}

\begin{figure}%
	\centering
	\begin{minipage}[b]{0.3\textwidth}
		\includegraphics[width=\textwidth]{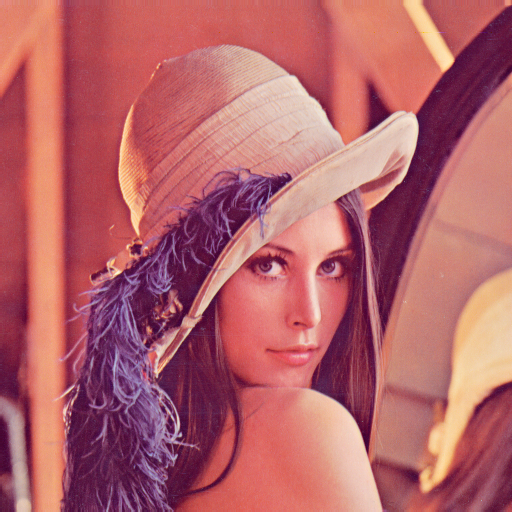}
		\caption{Plain Image}
	\end{minipage}\label{fig-4}
	\hfill
	\begin{minipage}[b]{0.3\textwidth}
		\includegraphics[width=\textwidth]{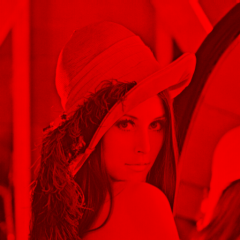}
		\caption{Red Channel}
	\end{minipage}\label{fig-5}
	\hfill
	\centering
	\begin{minipage}[b]{0.3\textwidth}
		\includegraphics[width=\textwidth]{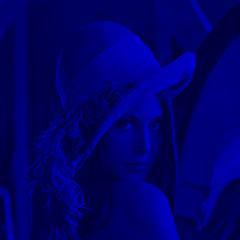}
		\caption{Blue Channel}
	\end{minipage}\label{fig-6}
\end{figure}

\begin{figure}%
	\centering
	\begin{minipage}[b]{0.3\textwidth}
		\includegraphics[width=\textwidth]{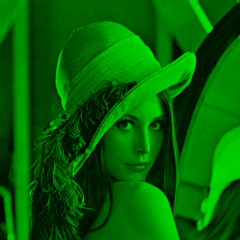}
		\caption{Green Channel}
	\end{minipage}\label{fig-7}
	\hfill
	\begin{minipage}[b]{0.3\textwidth}
		\includegraphics[width=\textwidth]{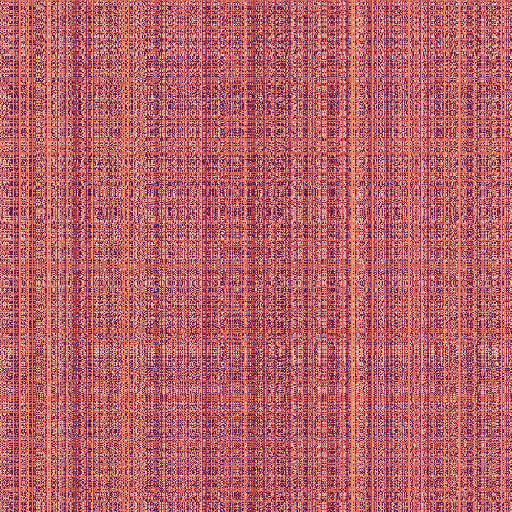}
		\caption{Swapped Pixels}
	\end{minipage}\label{fig-8}
	\hfill
	\begin{minipage}[b]{0.3\textwidth}
		\includegraphics[width=\textwidth]{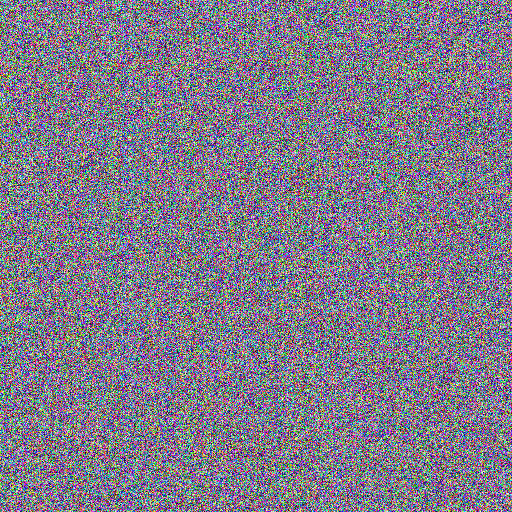}
	\caption{Cipher Image}
	\end{minipage}\label{fig-9}
\end{figure}

In addition to Xorshift, there is another PRNG, which is used by PiouCrypt: Triple Linear Congruential Generator (TLCG)); this component's task is to help for producing Bravais Lattice. An important point here is that TLCG is a PRNG, which runs three different linear congruential operations, and finally adds them together. On the other hand, the Triple Linear Congruential Generator proposed by Konsam and Loukrakpam (Tri-LCG), has some difference from TLCG the most important of which is the efficiency focus of Tri-LCG on hardware architecture, whereas TLCG is based on software side. Like all other Linear Congruential Generators, foundation of this generator is shown in equation \ref{eq-5}, and figure \ref{fig-10} shows a pseudo-code of TLCG.

\begin{align}
	x_{i+1} = ax_i + b \, \, (mod \, \, m)
	\label{eq-5}
\end{align}

\begin{figure}[h]%
	\centering
	\begin{program}
		\BEGIN \\ %
		\rcomment{m: Modulus (m\textgreater0), a: Multiplier (m\textgreater0), c: Increment (m\textgreater c$\geq$0), x: Seed (m\textgreater x$\geq$0)}
		\PROC |LCG\_1|(modulus, multiplier, increment, seed) \BODY
			\FOR true \DO
				seed := (multiplier * seed * increment);
				|return| \, \, seed; \OD \ENDPROC
		\PROC |LCG\_2|(modulus, multiplier, increment, seed) \BODY
			\FOR true \DO
				seed := (multiplier * seed * increment);
				|return| \, \, seed; \OD \ENDPROC
		
		\PROC |LCG\_3|(modulus, multiplier, increment, seed) \BODY
			\FOR true \DO
				seed := (multiplier * seed * increment);
				|return| \, \, seed; \OD \ENDPROC
		
		\PROC |TLCG|(modulus, multiplier, increment, seed) \BODY
			range := max - min;
			\rcomment{The values of \textit{seed}s are individual for every LCG in PiouCrypt}
			|print| (LCG\_1(m, a, c, x)+LCG\_2(m, a, c, x) \\ +LCG\_3(m, a, c, x) \, \, |mod| \, \, range) + min\ENDPROC
		\END
	\end{program}
	\caption{Pseudo-code of TLCG used in PiouCrypt}\label{fig-10}
\end{figure}

\begin{figure}%
	\centering
	\begin{minipage}[b]{0.65\textwidth}
		\includegraphics[width=\textwidth]{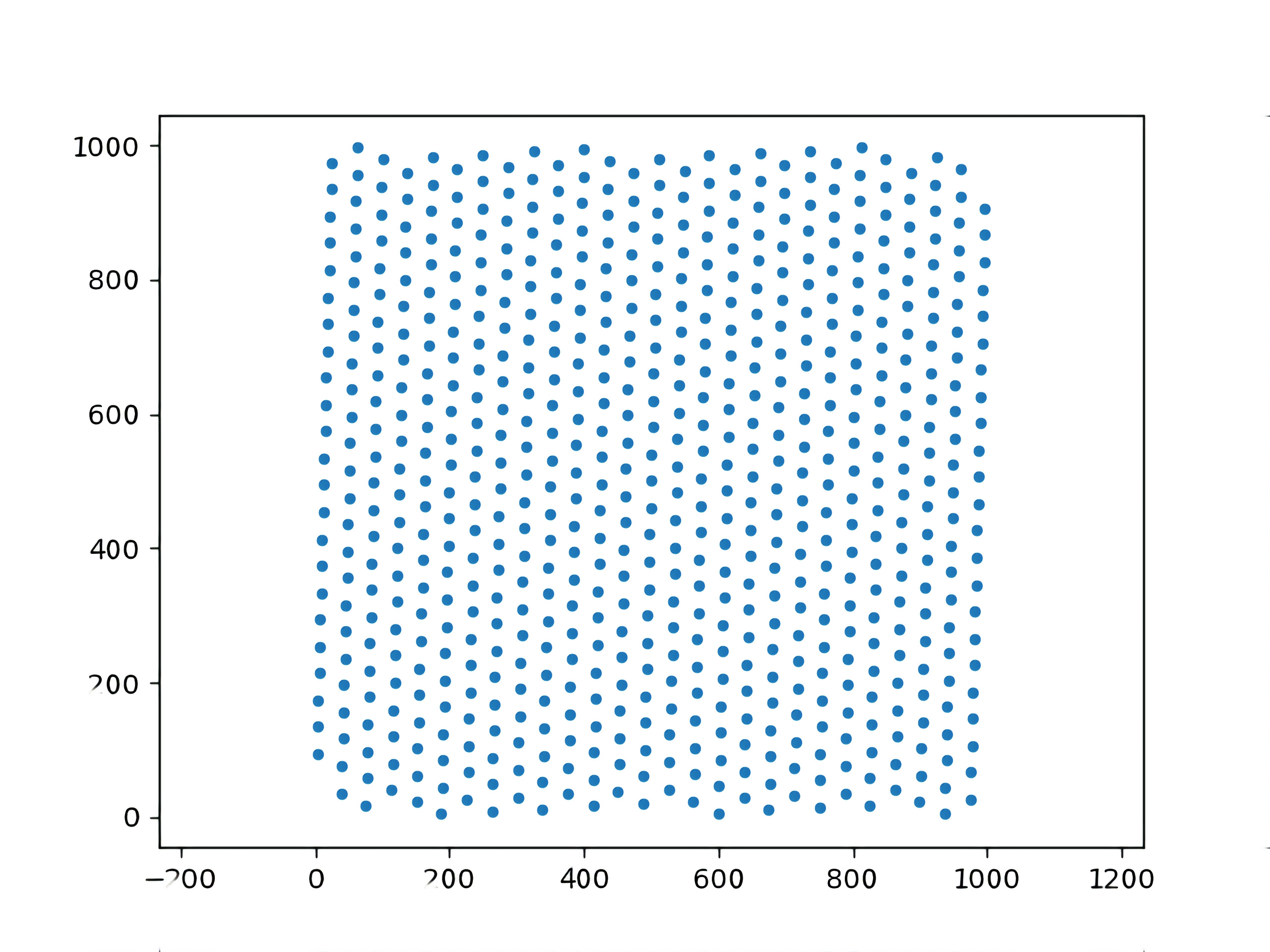}
	\end{minipage}
	\hspace*{\fill}
	\begin{minipage}[b]{0.3\textwidth}
		\includegraphics[width=\textwidth]{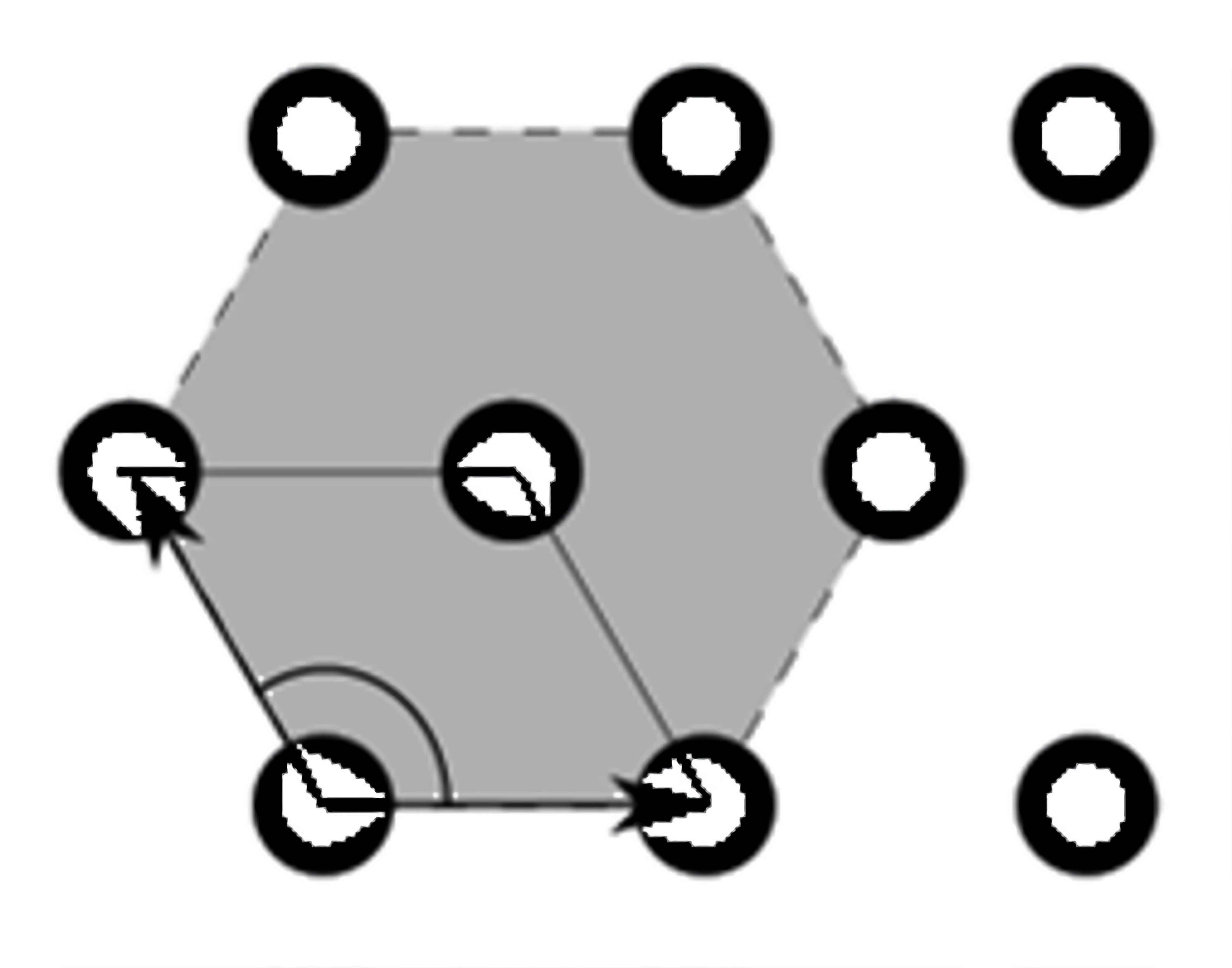}
	\end{minipage}
	\caption{Example of a Bravias hexagonal lattice produced in PiouCrypt with the size of 1000$\times$1000, and with Euclidean vectors $\hat{v}_{0}=(-40, -1)$ and $\hat{v}_{1}=(18,-37)$}
	\label{fig-11}
\end{figure}

After encrypting the plain image, a lattice structure based on Bravais Lattice Crystal System is going to get created by TLCG. Sometimes in physical sciences (specifically in Crystallography and Material Engineering) the details of lattice's unit cells can be shown similar to equation \ref{eq-6}:
\begin{align}
	A(\omega)=1-\mid S_{1}(\omega)\mid ^{2}-\mid S_{2}(\omega)\mid ^{2}
	\label{eq-6}
\end{align}
where $A(\omega)$, $S_{1}$, and $S_{2}$ are representing molar attenuation coefficient, reflected and transmitted powers \cite{bib34, bib35}. However, here in PiouCrypt from mathematical aspect the Bravias lattice is going to be a theoretical model, and considered like this:
\begin{align}
	\vec{r}_{n_{1} n_{2}}=n_{1} \hat{v}_{0}+n_{2} \hat{v}_{1}
	\label{eq-7}
\end{align}
In this 2D model, which is based on Bravais structure, the lattice $\vec{r}_{n_{1} n_{2}}$ has primitive translation vectors $\hat{v}_{0}$ and $\hat{v}_{1}$, and $n_{1}$ and $n_{2}$ as lattice point's indices. As it is tangible in figure \ref{fig-11}, the Bravias lattice of PiouCrypt is more like a hexagonal structure. So, the canonical basis vectors of lattice unit cells $\hat{c}_{0}$ and $\hat{c}_{1}$ can be shown as
\begin{align}
	\gamma_{1}\left(\cos \theta \hat{c}_{0}+\sin \theta \hat{c}_{1}\right)=\gamma_{0} \hat{c}_{0}, \hat{v}_{1}=\hat{v}_{0}
	\nonumber
	\\
	s.t. \quad \frac{\pi}{2} \geq \theta > 0
	\label{eq-8}
\end{align}
where $\theta$ represents obliquity, and $\gamma$ shows size of the canonical vectors (in hexagonal lattices $\gamma_{1}=\gamma_{0}$). After entering Security Layer 1 to Bravias System, two vectors will get generated based on height and width of the plain image by TLCG, and the Bravias lattice will get produced according to these vectors.

The next step (Non-negative Matrix Factorization) is to extract the points generated in Bravais Crystal System, and assign them into a 2D matrix $\mathbf{V}\in \mathbb{R}^{m \times n}$, then it will get factorized into two matrices $\mathbf{W}\in \mathbb{R}^{m \times r}_{+}$ and $\mathbf{H}\in \mathbb{R}^{r \times n}_{+}$, in a way that it can get re-approximated ($\mathbf{V} \cong \mathbf{WH}$). When $\mathbf{W}$ and $\mathbf{H}$ get calculated, the value of $\mathbf{W}$ is going to get written in \textit{OEA-key.txt} as the key for Security Layer 2. PiouCrypt is using Multiplicative Weight Update (Lee and Seung \cite{bib36-1}) Method for approximation (Algorithm \ref{algo-2}). It is noteworthy that Lee and Seung's method can perform more efficiently by adding some convolution (CNMF method) as equation \ref{eq-9}, where the values of $\mathbf{W}$ and $\mathbf{H}$ get computed by sufficiently small values $\alpha$ and $\beta$ (over\-fitting reduction parameters), $\omega=\sqrt{k n}-(\sqrt{k n}-1) \varGamma$ (s.t. $n$, $\varGamma$, and $k$ are showing column's number, sparsity, and rank of matrix respectively), and $\mathbf{C}$ represents problem-based variable constraints ($\mathbf{C}(\mathbf{W})=\|\mathbf{W}\|_{F}^{2}$); finally, $h$ is the product of mixing ratio and saturated mixing ratio \cite{bib36-2, bib37}. The output of factorization for lattice in fig. \ref{fig-11} is mentioned in equation \ref{eq-10}.

\begin{align}
	\mathbf{W}_{i j}^{(t-1)}&=(\alpha \frac{\partial J_{1}(\mathbf{W})}{\partial w_{i j}}+\left(\mathbf{W}^{(t-1)} \mathbf{H H}^{T}\right)_{i j})^{-1} \cdot
	(\left(\mathbf{V H}^{T}\right)_{i j}) \cdot 
	\mathbf{W}_{i j}^{(t)}
	\nonumber
	\\
	\mathbf{H}_{i j}^{(t-1)}&=(\beta \frac{\partial J_{2}(\mathbf{H})}{\partial h_{i j}}+\left(\mathbf{W}^{T} \mathbf{W} \mathbf{H}^{(t-1)}\right)_{i j})^{-1} \cdot
	(\left(\mathbf{W}^{T} \mathbf{V}\right)_{i j}) \cdot
	\mathbf{H}_{i j}^{(t)}
	\label{eq-9}
\end{align}

\begin{algorithm}
	\caption{Multiplicative Procedure for NMF}\label{algo-2}
	\begin{algorithmic}[1]
	\Require Matrix $V$
	\Ensure Matrices $W$ and $H$
	\State i $\Leftarrow$ 0
	\While{Maximum Number of Iterations $\geq$ i}
		\State $ H \Leftarrow H \times (W^{T}A)
		/ (W^{T}WH + 10^{-9})$
		\State $ W \Leftarrow W \times (AH^{T})
		/ (WHH^{T} + 10^{-9})$
		\State i $\Leftarrow$ i + 1
	\EndWhile
	\end{algorithmic}
\end{algorithm}

\begin{align}
	\mathbf{H}_{2 \times 2} =
	\begin{bmatrix}
		2.67699 & 6.99999 \\
		8.69999 & 4.47100
	\end{bmatrix}
	\quad \quad
	\mathbf{W}_{668 \times 2} =
	\begin{bmatrix}
		3.53299 & 2.09400 \\
		3.60300	& 2.01000 \\
		\quad \quad \quad \quad \vdots \\
		6.19999 & 1.68999 \\
		1.32000 & 8.49999
	\end{bmatrix}
	\label{eq-10}
\end{align}

\begin{figure}%
	\centering
	\begin{minipage}[b]{0.45\textwidth}
		\includegraphics[width=\textwidth]{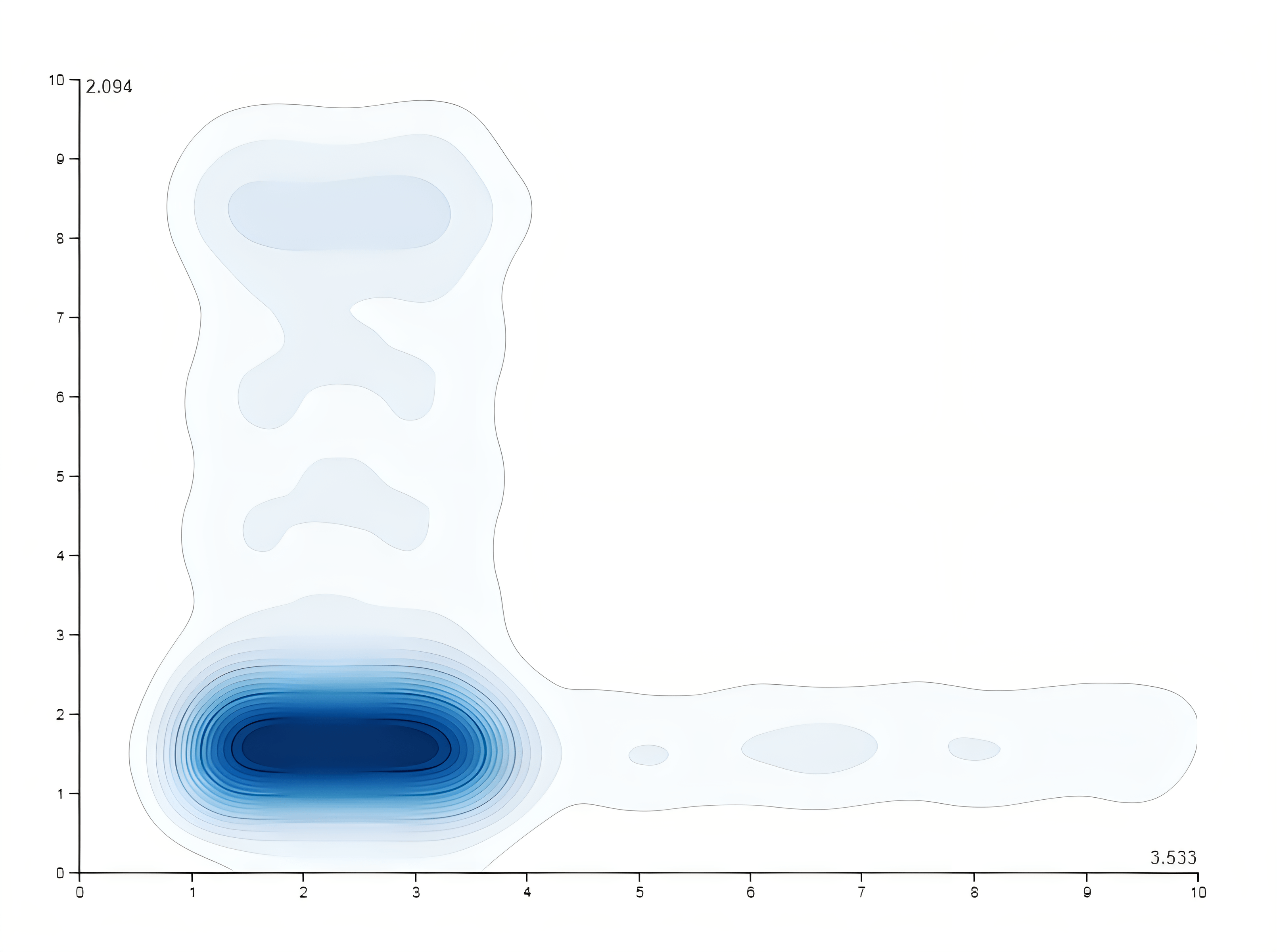}
	\end{minipage}
	\hspace*{\fill}
	\begin{minipage}[b]{0.45\textwidth}
		\includegraphics[width=\textwidth]{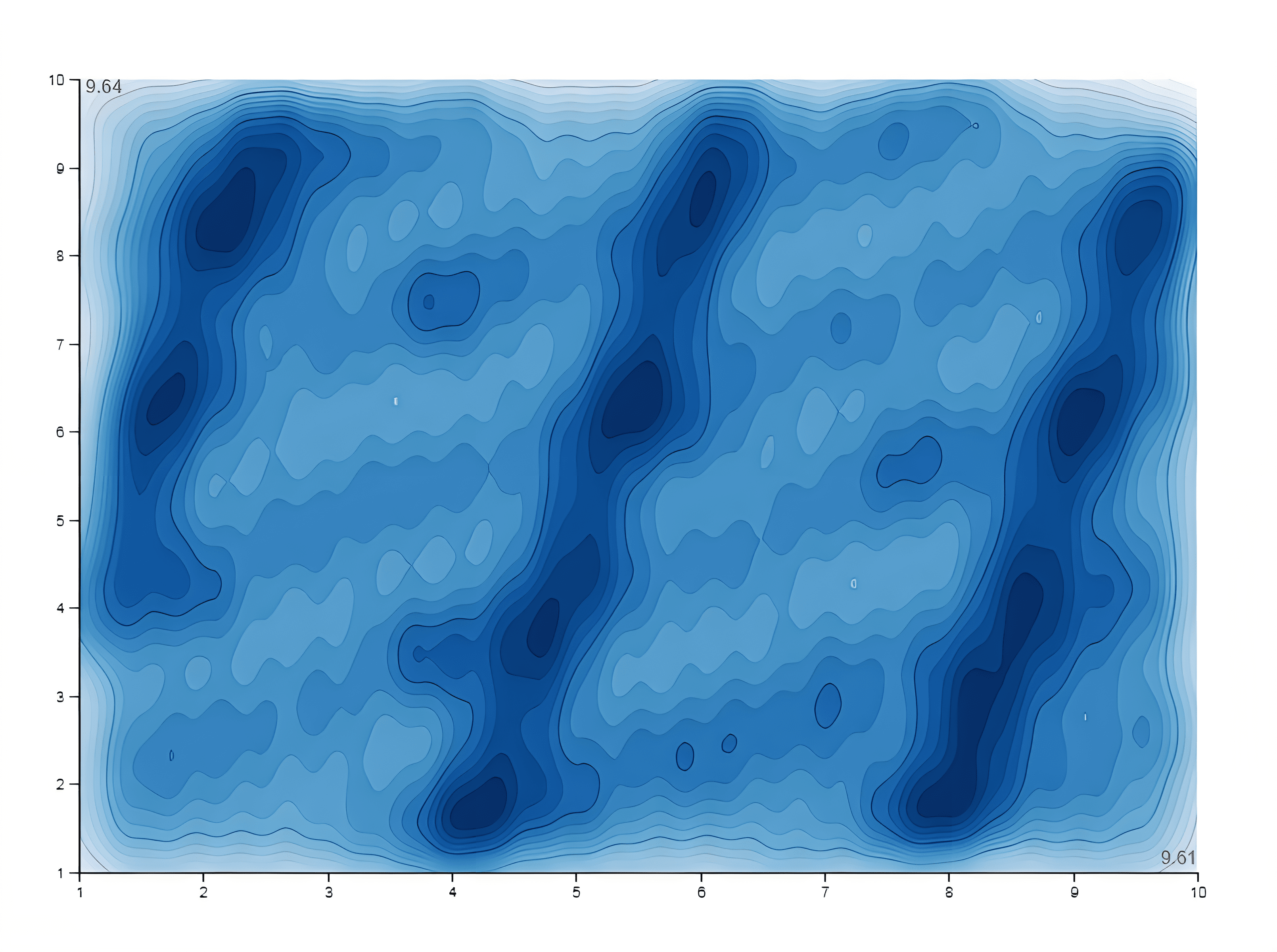}
	\end{minipage}
	\caption{Isopleths of lattice's points as $\mathbf{V}$ in fig. \ref{fig-11} (right figure), and it's factorized value as the matrix $\mathbf{W}$ (left figure)  (legend: the darker color shows higher density of values)}
	\label{fig-12}
\end{figure}

\subsection{Secondary Security Layer}\label{subsec3_2}
The next security level, which is based on Odd-Even Attribute (OEA), is a novel and easy-to-compute technique based on parity attribute and ASCII encoding, which can get executed on most of the modern computational systems and telecommunications equipment. Most of the other encoding standards are compatible with ASCII, for example, UTF-8 is one of the famous encoding methods at the web space which is correspondent with ASCII's standard. By using ASCII, computers can only store numbers, not letters (characters); so, everything has to be represented by numbers, either a single number or a set of multiple numbers. 

\begin{figure}[h]%
	\centering
	\includegraphics[trim={0 0 1mm 0},clip, width=1\textwidth]{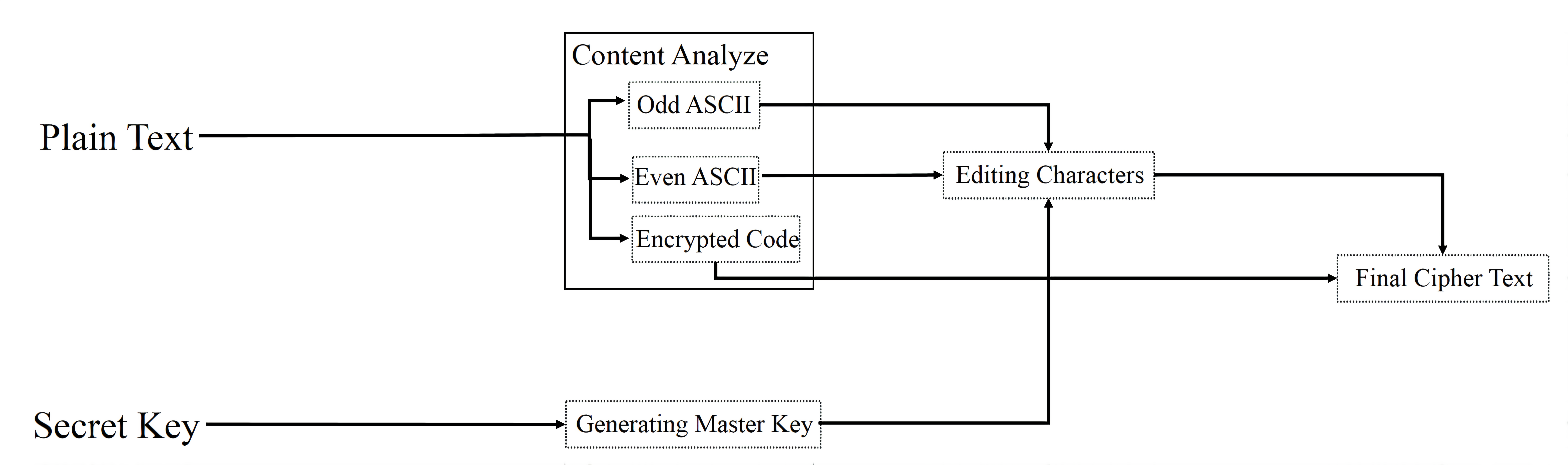}
	\caption{Procedure of Secondary Security Layer}\label{fig-13}
\end{figure}

Detailed function of Security Layer 2 is expressed in algorithm \ref{algo-3}, but OEA has several small components that information flows among them during process:
\begin{itemize}
	\item \textit{Weight}: consists of summing a component's ASCII codes
	\item \textit{Secret Key}: contents of \textit{OEA-key.txt}, which is going to get used for encrypting the data
	\item \textit{Master Key}: (length of plain text)$\times$(secret key’s weight) 
	\item \textit{Odd String (OS)}: repository to store characters that have odd ASCII code
	\item \textit{Even String (ES)}: a storage for the characters with even ASCII code
	\item \textit{Encrypted String (ENS)}: characters that have been encrypted by \textit{0} and \textit{1}
	\item \textit{Redundant Strings (red\_str)}: portions to create some diffusion in cipher text
\end{itemize}

\begin{algorithm}
	\caption{OEA Encryption}\label{algo-3}
	\begin{algorithmic}[1]
		\Require (plain text: \textit{Layer\_1-keys.txt}, key: \textit{OED-key.txt})
		\Ensure cipher text consists of properties of Security Layer 1 (key for Bob)
		\State \textbf{transform} secret\_key \textbf{to} ASCII\_Code, i $\Leftarrow 0$
		\While{i $<$ size of secret key}
			\State $secret\_key\_weight \Leftarrow secret\_key\_weight + secret\_key_{i}$
			\State i $\Leftarrow$ i $+ 1$
		\EndWhile
		\State $master \, \, key \Leftarrow secret\_key\_weight \, \times$ size of plain text, i $\Leftarrow 0$
		\While{i $<$ size of plain text}
			\If{i (mod 2) $\neq$ 0}
				\State $SO_{i} \Leftarrow plaintext_{i}$, \, $SC_{i} \Leftarrow "1"$
			\Else
				\State $SE_{i} \Leftarrow plaintext_{i}$, \, $SC_{i} \Leftarrow "0"$
			\EndIf
		\State i $\Leftarrow$ i $+ 1$
		\EndWhile
		\State $SE_{0} = SE_{0} - master\_key$, i $\Leftarrow 0$
		\While{i $<$ size of \textit{SE}}	
		\State $SE_{i} \Leftarrow SE_{i} + SE_{i-1}$, i $\Leftarrow$ i $+ 1$
		\EndWhile
		\State $SE_{last \, element} \Leftarrow SE_{last \, element} - master\_key$
		\State $SO_{0} \Leftarrow SO_{0} - master\_key$, i $\Leftarrow 0$
		\While{i $<$ size of \textit{SO}}	
		\State $SO_{i} \Leftarrow SO_{i} + SO_{i-1}$
		\State i $\Leftarrow$ i $+ 1$
		\EndWhile
		\State $SO_{last \, element} \Leftarrow SO_{last \, element} + master\_key$
		\State size of $red\_str1$ and $red\_str2$ $\Leftarrow$ weight of secret key (mod 10)
		\State i,j $\Leftarrow 0$
		\While{i $<$ size of \textit{red\_str1}}	
			\If{$secret\_key_{i} (mod 2) = 0$}
				\State $red\_str1_{i} = "1"$
			\Else
				\State $red\_str1_{i} = "0"$
			\EndIf
			\If{j = size of $secret\_key$}
				\State j $\Leftarrow$ 0
			\EndIf
			\State i $\Leftarrow$ i $+ 1$, j $\Leftarrow$ j $+ 1$
		\EndWhile
		\State i $\Leftarrow 0$
		\While{i $<$ size of \textit{red\_str2}}	
			\State $red\_str2_{i} \Leftarrow secret\_key_{i} + master\_key$
			\If{$i$ = size of $secret\_key$}
				\State i $\Leftarrow$ 0
			\EndIf
			\State i $\Leftarrow$ i $+ 1$
		\EndWhile	
	\end{algorithmic}
\end{algorithm}

After entering inputs into second security layer's encryption algorithm, the OEA starts to transform the secret key string into an ASCII code integer variable named \textit{ASCII\_Code}; in fact, this part of algorithm puts every secret key's characters ASCII code into a particular variable. Then, the length of the plain text, and weight of the secret key will be calculated. Afterwards, while traversing on the plain text, the algorithm will check the odd or even attribute of character's ASCII's code:
\begin{itemize}
	\item By facing with odd code, that character will be added into \textit{SO}, and number \textit{1} will be subjoined into \textit{SC}.
	\item Otherwise, the algorithm will place intended character in the \textit{SE}, and number \textit{0} is going to be appended to \textit{SC}.
\end{itemize}

OEA tries to add huge intricacy by putting small operation together:
\begin{itemize}
	\item First of all, the \textit{SE} will get traversed and in value of master key will subtracted from first and last character's ASCII code, and during traversing on \textit{SE}, value of every character's ASCII code will be summed by the previous character's code. Thereupon, it's \textit{SO} string's turn to get traversed and OEA will add the value of master key to first and last character of \textit{SO}, and while traversing on \textit{SO}, value of every character's ASCII code will be subtracted by the previous character's code.
	\item Next, it's time to append two redundant strings into the beginning of \textit{SC} and end of \textit{SO}. This new string's size will equal to module of secret key's weight on \textit{10}.
	\item Then, the algorithm starts traversing on the secret key and it fills the content of $red\_strl$ with number $0$ or $1$ due to each character's ASCII code; on the other hand: $red\_str2 = Secret\_key_{i} + master\_key + i$.
	\item The point is that, if the algorithm reached to end of the secret key, it will begin again from start point until $red\_strl$ and $red\_str2$ are filled.
\end{itemize}
To wrap up and present the output, processed information for creating the key for Bob will export respectively in this order: \textit{red\_str1, SC, SE, SO, red\_str2}.
\\
So, the encrypted image can get delivered to Bob, along with the key generated in Security Layer 2 (decryption process is obvious for both first and second layer).

\section{Processing Image Against Adversary}\label{sec4}
Most of the common visual cryptography techniques are depended on one layer of operations, but investigating and analyzing the security of a method similar to PiouCrypt may need to deeper look at details from visual aspects. \par{}
\begin{figure}
	\centering
	\includegraphics[trim={0 0 1mm 0},clip, width=0.4\textwidth]{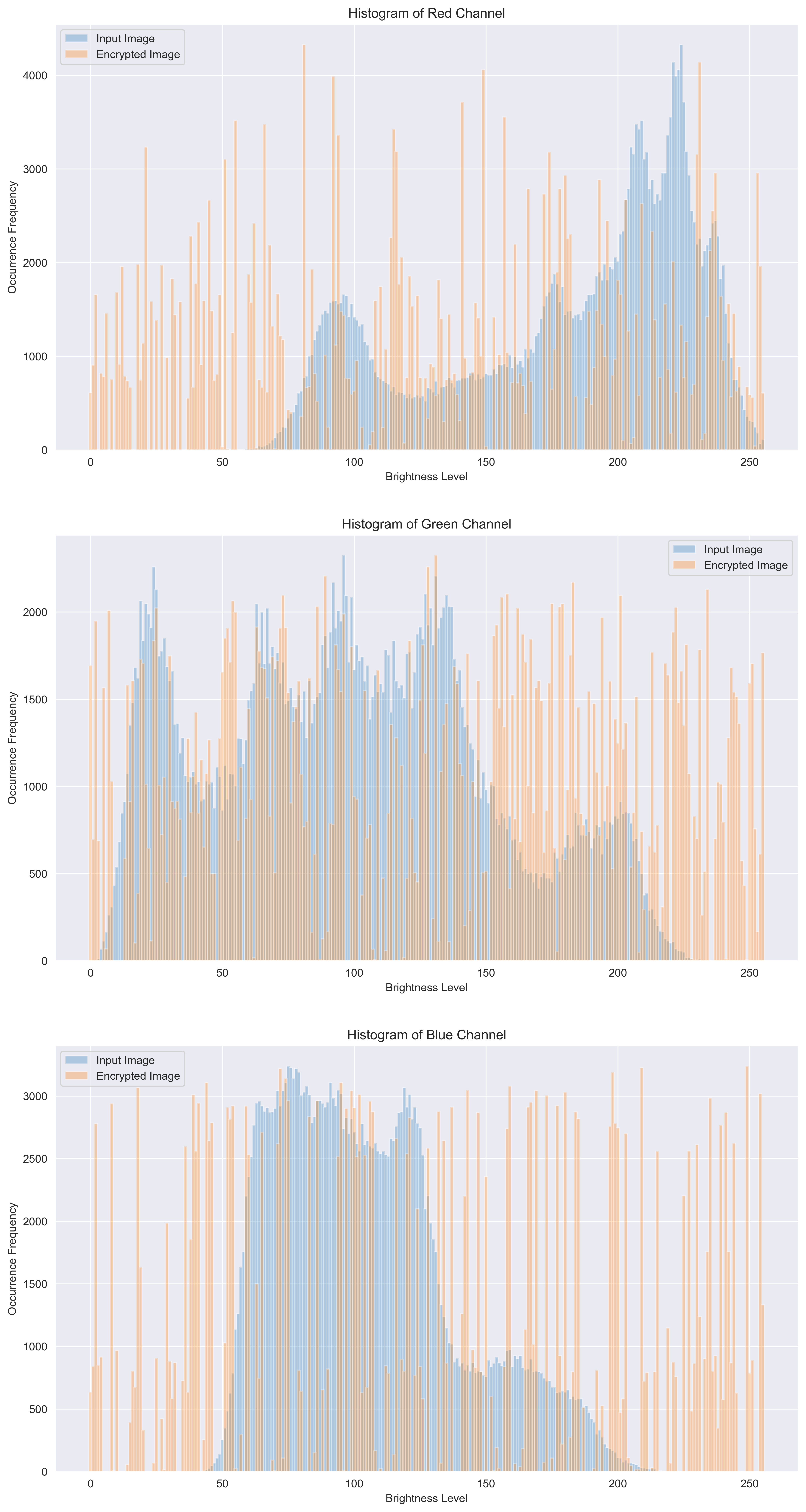}
	\caption{Histogram Analysis for PiouCrypt Method}\label{fig-14}
\end{figure}
The histogram on brightness levels on each present channel in a digital image is a non-semantic analysis that stands in the point processing phase of digital image topics where it provides distribution information on a given channel and is often used for different image enhancement purposes. The histogram of a digital image with gray levels in the range $[0, L-1]$ is a discrete function $h(r_{k}) = n_{k}$ where $r_{k}$ is the $k^{th}$ gray level, $n_{k}$ is the number of pixels in the image having gray level $r_{k}$ and $h(r_{k})$ is the histogram of target digital image. The Lenna's RGB image is used with 512$\times$512 dimensions and 24 bits depth as the input image which is shown in figure \ref{fig-4}. After applying our proposed method to the input image, the encrypted image was obtained as it is observable in figure \ref{fig-9} where it has the same dimensions and bit depth as the input image. Then, histogram analysis is performed on the present image, and the results are obtained as charts in figure \ref{fig-14}.

\section{Results and Comparisons}\label{sec5}

\begin{figure}%
	\centering
	\begin{minipage}[b]{0.3\textwidth}
		\includegraphics[width=\textwidth]{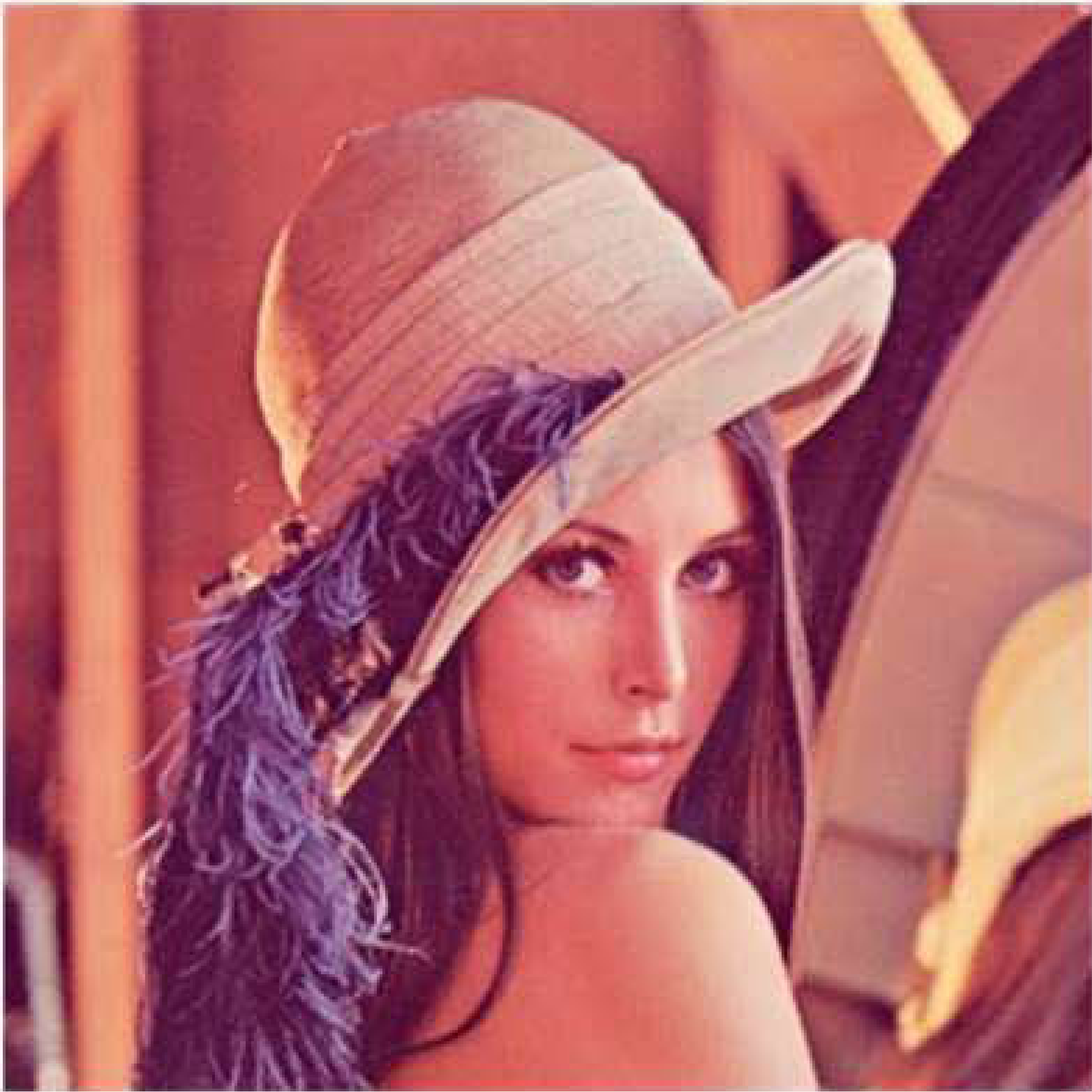}
		\caption{Ref. \cite{bib26}'s Plain Image}
	\end{minipage}\label{fig-17}
	\hfill
	\begin{minipage}[b]{0.3\textwidth}
		\includegraphics[width=\textwidth]{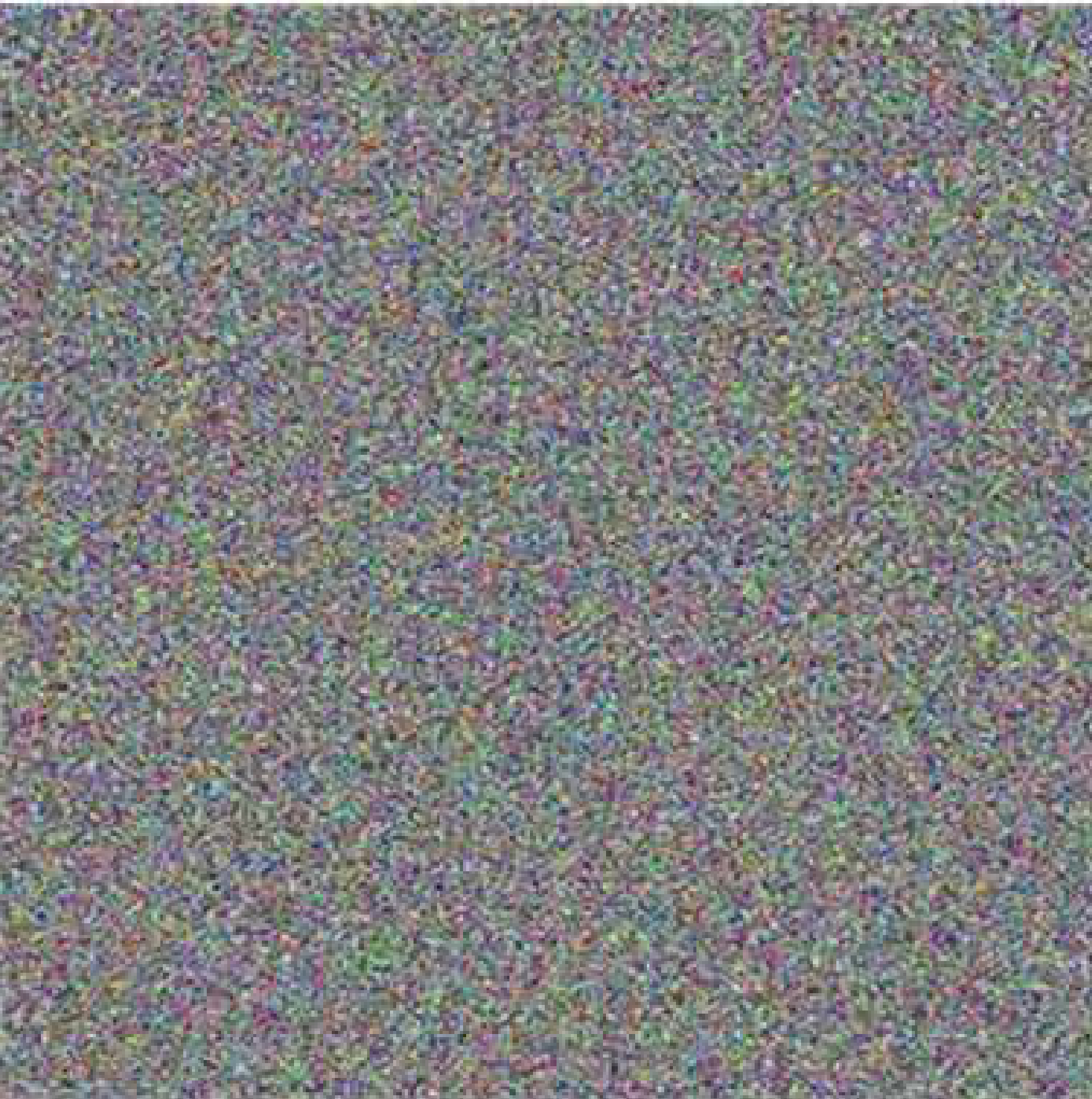}
		\caption{Ref. \cite{bib26}'s Encrypted Image}
	\end{minipage}\label{fig-16}
	\hfill
	\centering
	\begin{minipage}[b]{0.3\textwidth}
		\includegraphics[width=\textwidth]{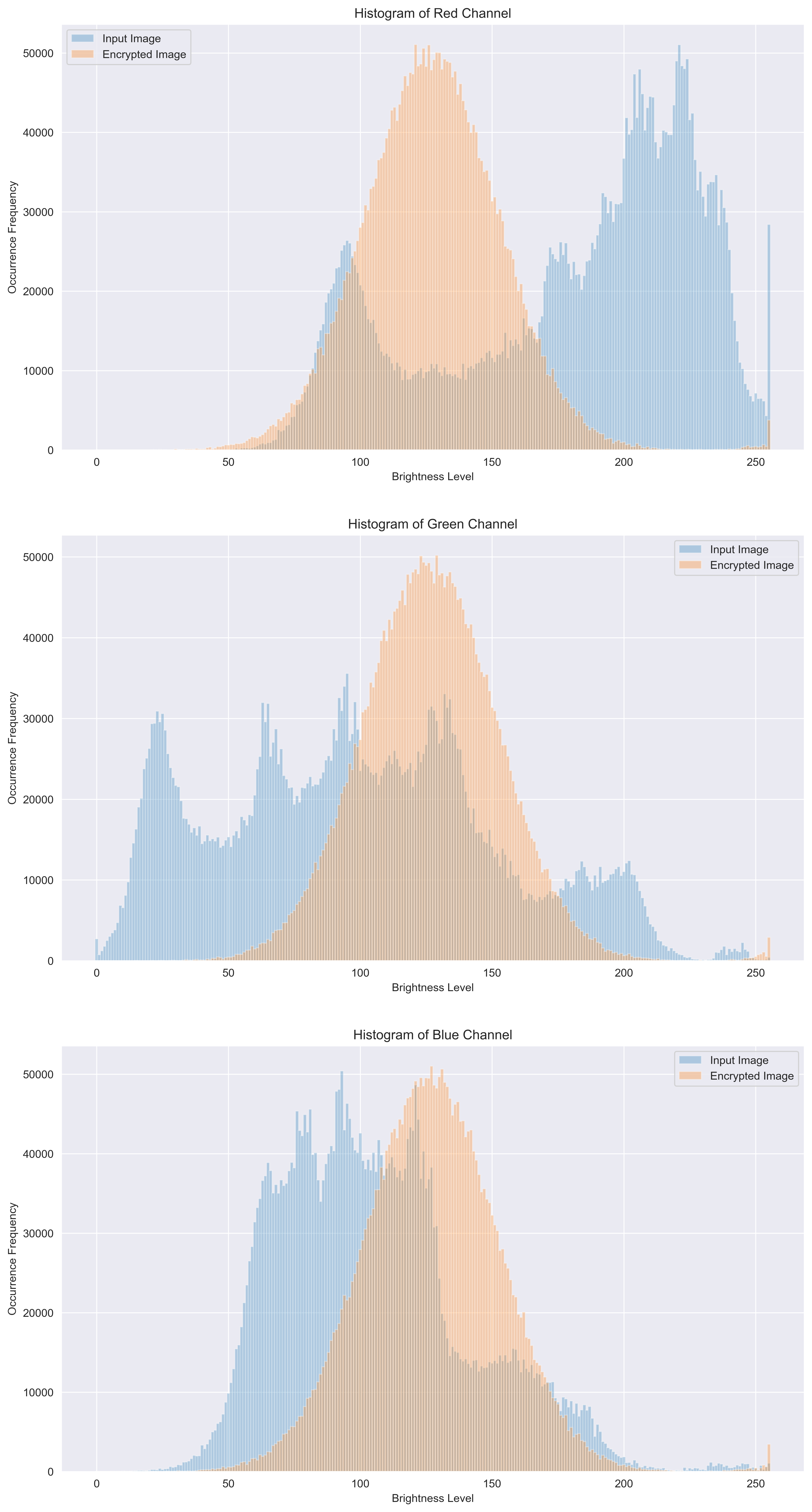}
		\caption{Ref. \cite{bib26}'s Histogram}
	\end{minipage}\label{fig-18}
\end{figure}

As indicated in charts of figure \ref{fig-18}, histogram of encrypted Lenna's RGB image using proposed method in \cite{bib26} has the Gaussian distribution with estimated mean $\sim127$ and standard deviation $\sim26$ same on triple channels. \par{}

\begin{figure}%
	\centering
	\begin{minipage}[b]{0.3\textwidth}
		\includegraphics[width=\textwidth]{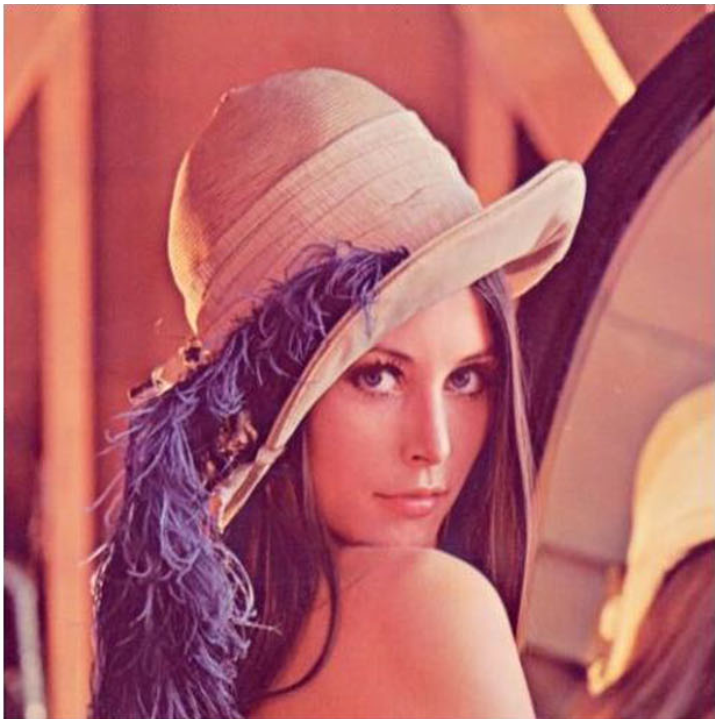}
		\caption{Ref. \cite{bib29}'s Plain Image}
	\end{minipage}\label{fig-19}
	\hfill
	\begin{minipage}[b]{0.3\textwidth}
		\includegraphics[width=\textwidth]{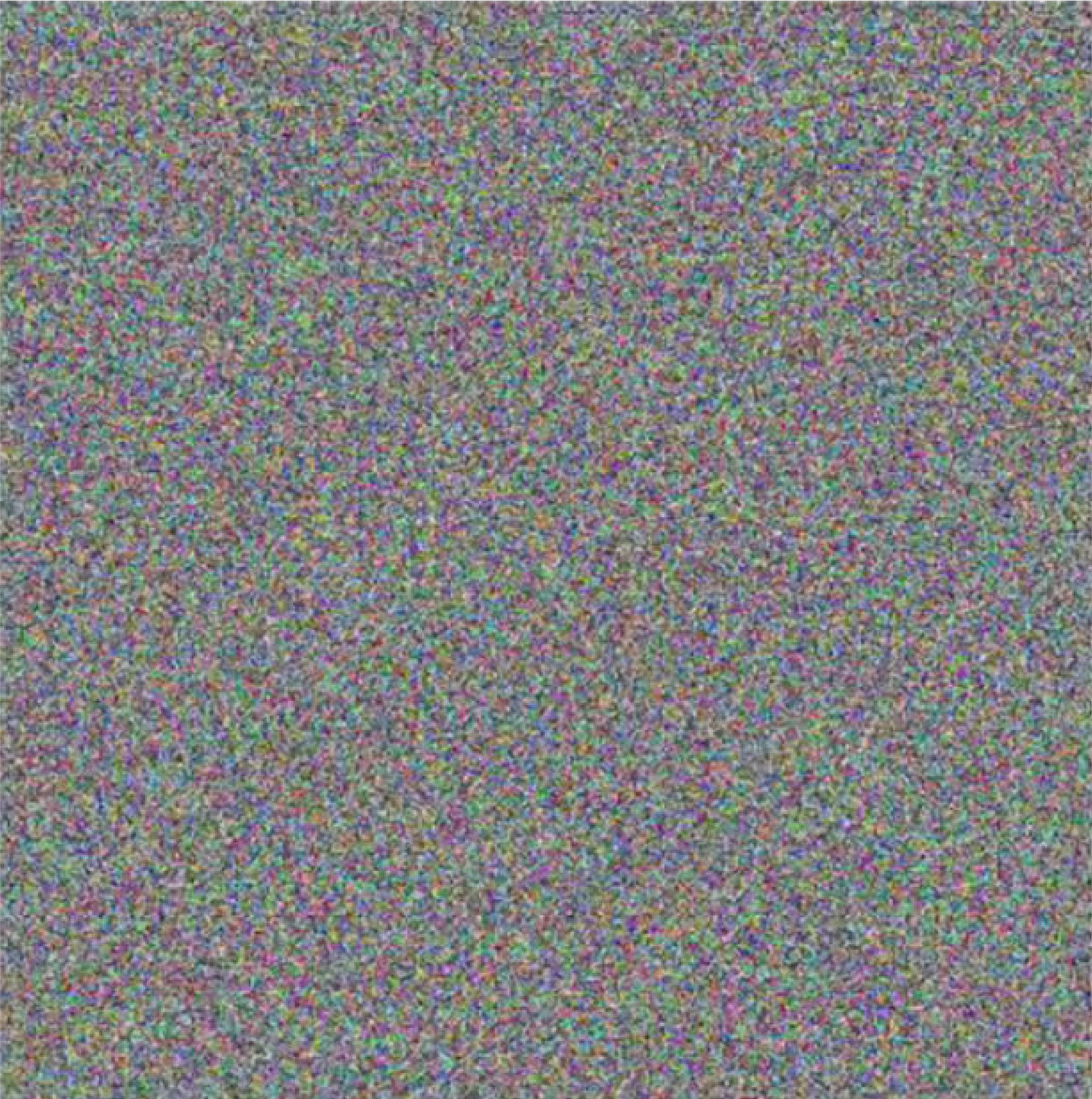}
		\caption{Ref. \cite{bib29}'s Encrypted Image}
	\end{minipage}\label{fig-20}
	\hfill
	\centering
	\begin{minipage}[b]{0.3\textwidth}
		\includegraphics[width=\textwidth]{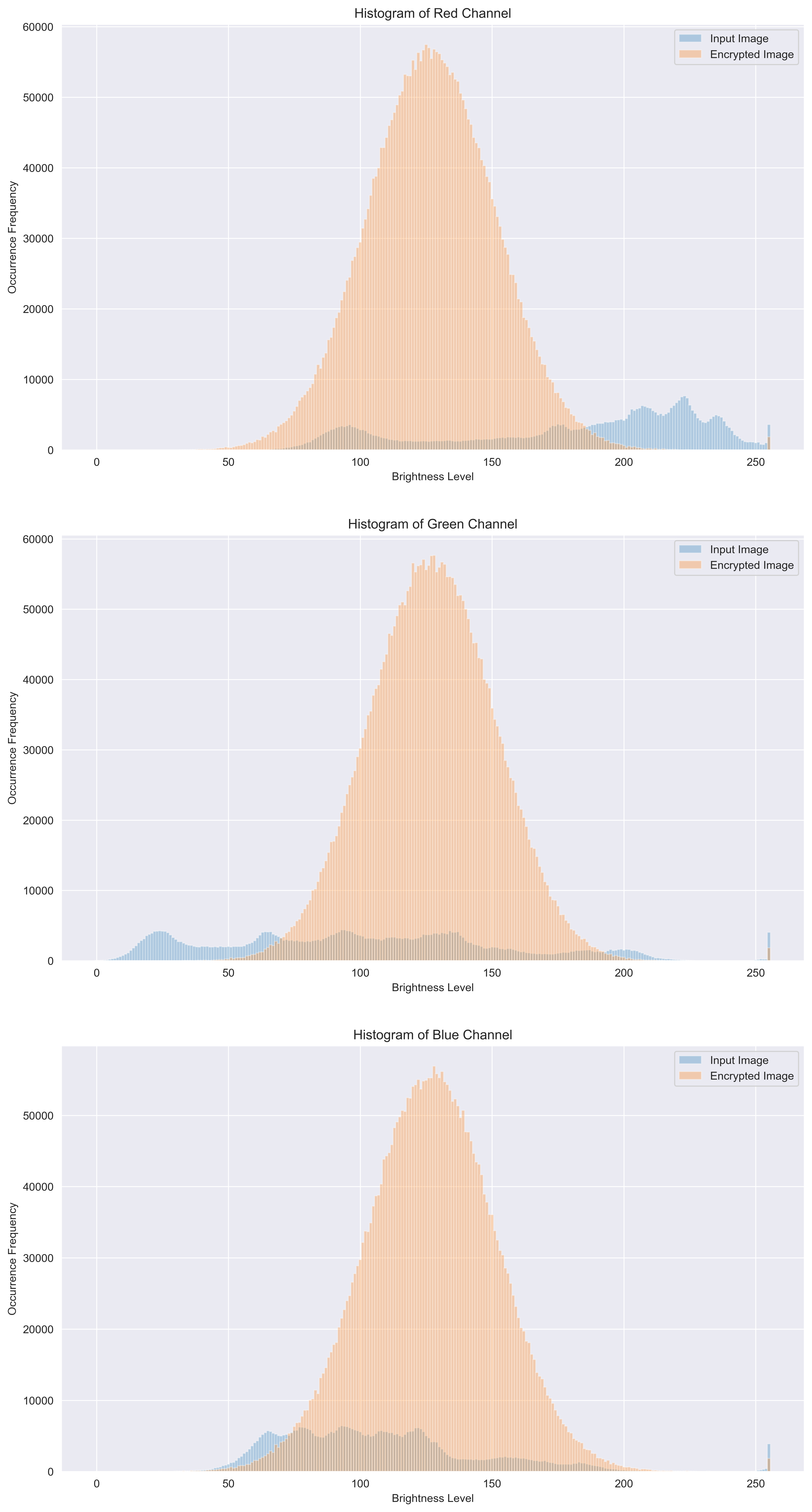}
		\caption{Ref. \cite{bib29}'s Histogram}
	\end{minipage}\label{fig-21}
\end{figure}

Also, according to charts of figure \ref{fig-21}, encrypting results for a given image to \cite{bib29} have the same Gaussian distribution but with a slight difference in parameters $(\sim127, \sim24)$ and smoothness where the second one has smoother and it is accurate shape to an exact normal distribution. In both methods, the positions of brightness levels in histogram of encrypted results are independent to input image brightness levels. \par{}

\begin{figure}%
	\centering
	\begin{minipage}[b]{0.3\textwidth}
		\includegraphics[width=\textwidth]{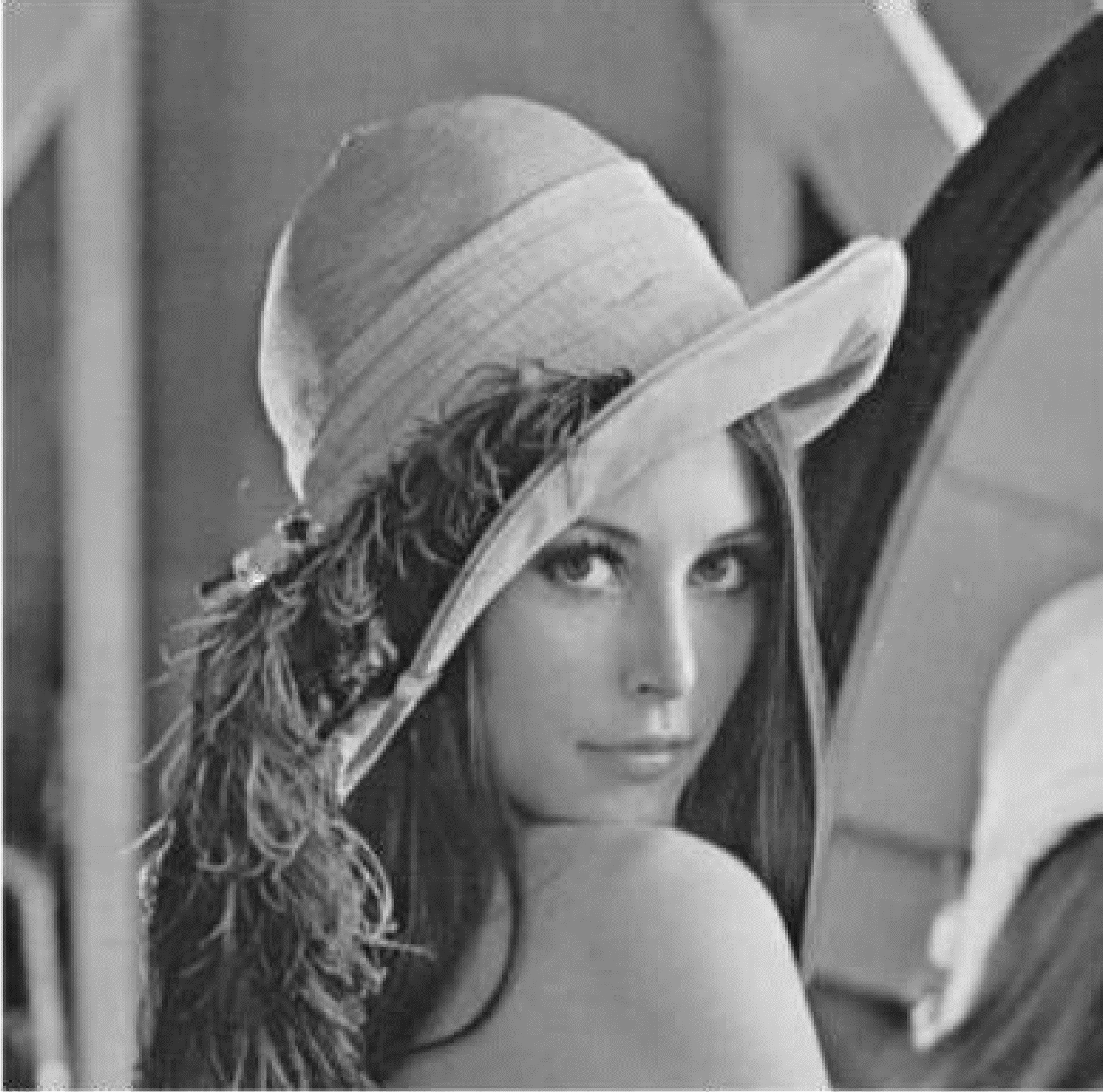}
		\caption{Ref. \cite{bib30}'s Plain Image}
	\end{minipage}\label{fig-22}
	\hfill
	\begin{minipage}[b]{0.3\textwidth}
		\includegraphics[width=\textwidth]{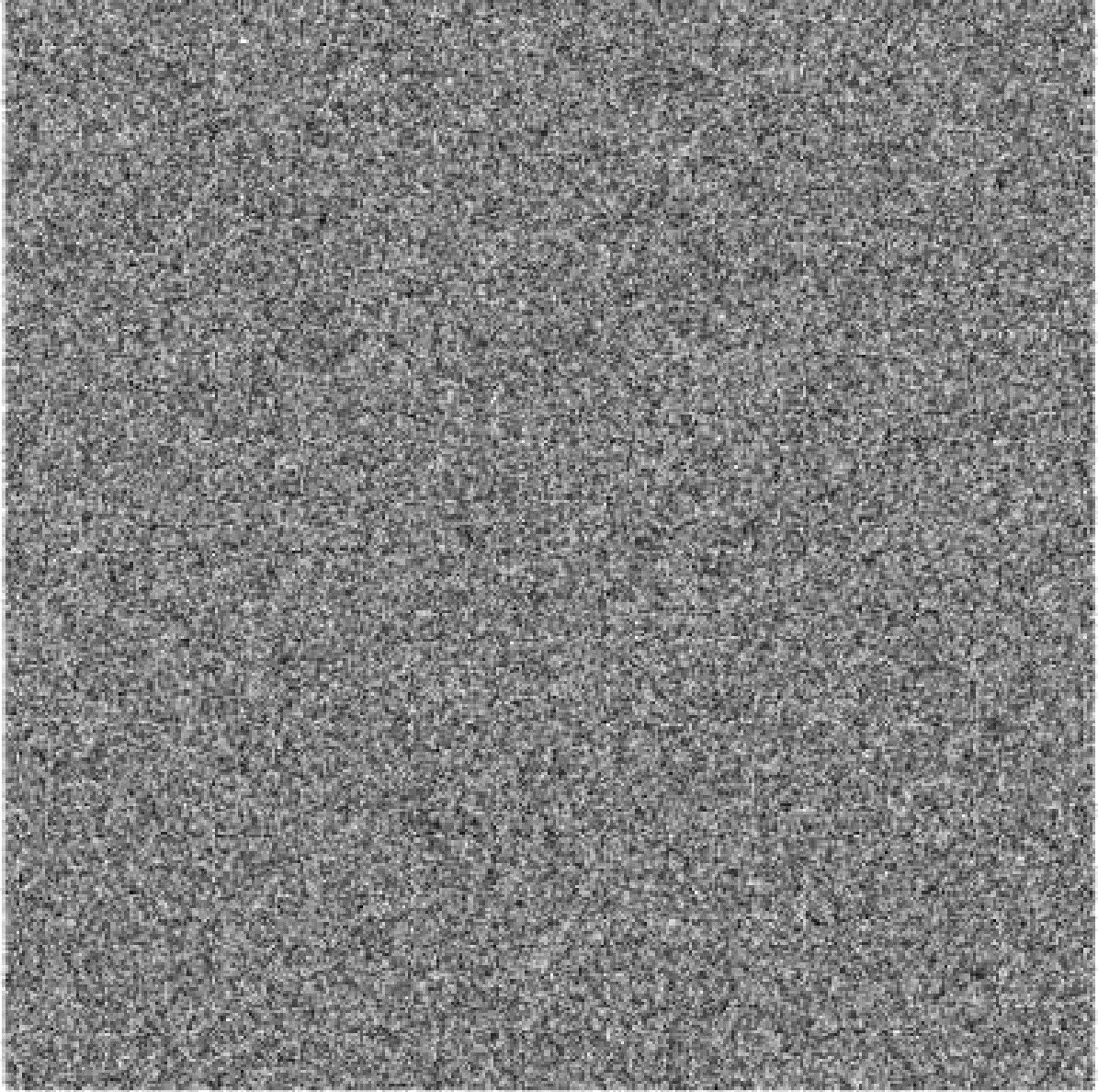}
		\caption{Ref. \cite{bib30}'s Encrypted Image}
	\end{minipage}\label{fig-23}
	\hfill
	\centering
	\begin{minipage}[b]{0.3\textwidth}
		\includegraphics[width=\textwidth]{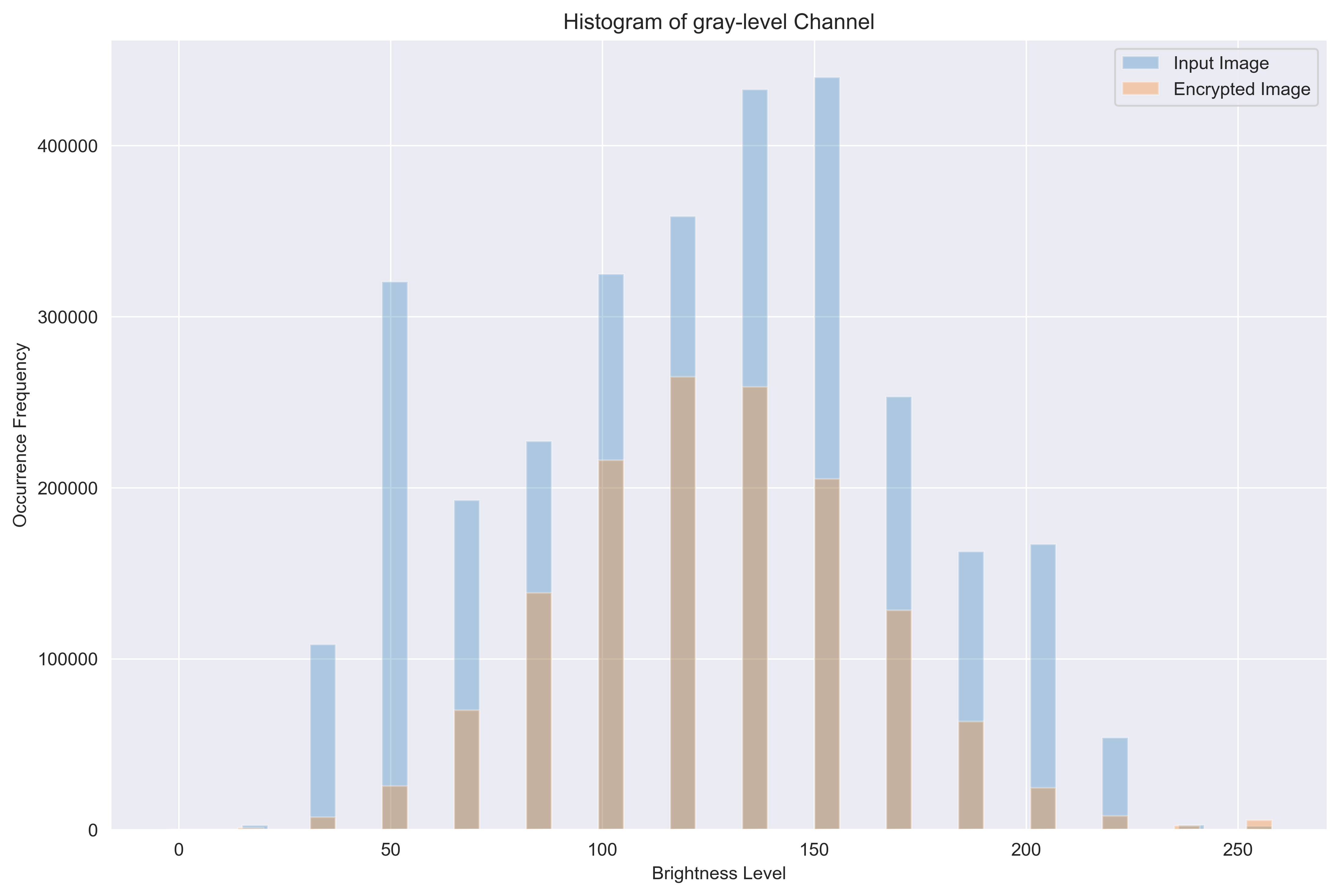}
		\caption{Ref. \cite{bib30}'s Histogram}
	\end{minipage}\label{fig-24}
\end{figure}

Figure \ref {fig-24} shows that \cite{bib30} which mostly worked on the single gray-level channel, the encrypted image has the same Gaussian distribution $(\sim127, \sim35)$. Despite of \cite{bib26} and \cite{bib29}, the positions of brightness levels in the histogram of encrypted results are dependent on input image brightness levels. \par{}

The results for PiouCrypt are showing that the encrypted image has no predefined distribution. Meanwhile, the result of each RGB channel is different from the others wherein each case was based on the dispersion of brightness levels in the input image. Also, brightness levels in the encrypted image are independent of the input image and despite of \cite{bib26}, \cite{bib29} and \cite{bib30} are stretched in the entire range $0-255$. In other words, these three methods are applying normal distribution to the channels of input image, but PiouCrypt circulating on random actions and shuffling. \par{}

Another thing in comparing PiouCrypt with \cite{bib26}, \cite{bib29} and \cite{bib30} is loss of data. The transformation of a non-normal distribution into a normal one is not going to be a injective and recursive function. Therefore, there is a transformation risk (even using maximum likelihood estimation) in the these related works, because from the point of histogram equalization, some of the data is losing and reverse function is not able to exactly map each pixels with original ones. On the other hand, PiouCrypt is using a lookup table instead of normal distribution, so no matter how much operations get complex, the transformation risk in PiouCrypt is $0$, and no data will loss during encryption/decryption. From the point of applicability, the best case for \cite{bib26}, \cite{bib29} and \cite{bib30} is recording the mapping data for preventing loss of data; although, the volume of stored information will markedly increase (e.g. the amount of saved records is going to equal with size of at least two pictures).

Eventually, about second security level in PiouCrypt, it should be mention that flexibility of OEA is effectively helpful; in fact, OEA has not any specific limitation for secret key, and it can get used separately by itself in other cryptographically systems. In addition to different types of diffusing and confusing for making adversary's work hard in Algorithm \ref{algo-3}, it can be used by the parties who have no pre-defined security arrangements for exchanging data, so usefulness of this algorithm at the encrypting huge amount of data can be tangible, apart form it's easily understandable algorithm.

\section{Discussion and Future Works}\label{sec6}
During theoretical creation process of PiouCrypt, incompatibility of OpenCV-python with several JPG and JPEG files, and inapplicability of multiple PRNGs were two things that attracted authors attention. \par{}

According to it's flexible structure, developers of PiouCrypt have planned to add these features in the future:

\begin{itemize}
	\item In order to investigate security factors deeper, PiouCrypt's components will get more entangled by turning it into a hybrid method for centralized universally composable protocols. Also, the encapsulation of data for hybrid structure will be in a way that the size of image get reduced, along with encrypting it.
	\item Generalizing the applicability of PiouCrypt for visual real-time communications is another part of future works. Keeping distance from heavy mathematical operations coupled with maintaining the balance of security and time is not enough, and when we are talking about real-time communication the efficiency of every component's algorithm is playing an outstanding role.
	\item Using a lattice component in the main part of PiouCrypt may give it potential for working effectively in post-quantum era, but it is not going to be enough, if more unconventional models of computation invent. So, instead of a NP-hard problem like non-negative matrix factorization, a problem should get used in a way that it need an algorithm to solve it in bounded-error quantum polynomial time.
\end{itemize}

\section{Conclusion}\label{sec7}
In this paper a novel technique for securing visual communication from the dimension of cryptography was presented entitled "PiouCrypt". This method is not centralized on one level of security and it is layered into two levels. At the outset, a picture enters into first security level, and PiouCrypt starts to encrypt it by permutations and shuffling the pixels. Then, a lattice structure gets created based on properties of plain image, and a non-negative matrix factorization is getting applied to it's data. Factorized outputs are operating as input key for second security level coupled with shuffling data as plain text. The second security level is a fast procedure for encrypting data in ASCII encoding. In the end, cipher text of second security level plays the role of key for the receiver, and encrypted image is going to send to him/her, along with the key. Investigating are showing the applicability of this method based on making the pixels of cipher image highly diffused for the attacker, and even histogram analysis of cipher image can not be so helpful for adversary because of shuffling columns and rows. The data related to swapping pixel sets are getting encrypted by a separated layer, so the adversary's chance for implementing different scenarios with powerful computational systems is highly ineffectual. 


\end{document}